\documentclass[floatfix,
prb,
superscriptaddress,
reprint,
preprint,
amsmath,
twocolumn,
amssymb,
amsfonts,
aps,10pt]{revtex4-2}

\usepackage{bm}
\usepackage[utf8]{inputenc}
\usepackage{etoolbox} 
\usepackage{enumerate}
\usepackage{enumitem}
\usepackage{graphicx}
\usepackage{mwe}
\usepackage{subfigure}
\usepackage{xcolor} 
\usepackage{array}
\usepackage{placeins}
\usepackage[nottoc]{tocbibind}
\usepackage{amssymb}
\usepackage{amsmath}
\usepackage{soul}
\usepackage[mathlines]{lineno}
\usepackage{appendix}
\usepackage[nottoc]{tocbibind}
\usepackage{xr}

\begin{document}

\title{Trigonal Distortion Driven Ground States in VX3 (X = Br and I)}

\author{Chamini S. Pathiraja}
\affiliation{Physics Department and Texas Center for Superconductivity, University of Houston, Houston, TX 77204}

\author{Deniz Wong}
\author{Christian Schulz}
\affiliation{Helmholtz-Zentrum Berlin für Materialien und Energie, D-14109 Berlin, Germany}

\author{Yi-De Chuang}
\affiliation{Lawrence Berkeley National Laboratory, Berkeley, CA 94720}

\author{Yu-Cheng Shao}
\affiliation{National Synchrotron Radiation Research Center,  101 Hsin-Ann Road, Hsinchu Science Park, Hsinchu, Taiwan 30076}

\author{Byron Freelon}
\email[Corresponding author:]{bkfreelo@central.uh.edu}
\affiliation{Physics Department and Texas Center for Superconductivity, University of Houston, Houston, TX 77204}

\date{\today}

\begin{abstract}

Transition-metal halides V$X_3$ ($X$ = Br and I) have emerged as promising two-dimensional magnetic materials for future spintronic applications, yet their ground state electronic properties remain poorly understood. Here, we employ high-resolution resonant inelastic x-ray scattering (RIXS) combined with ligand-field multiplet calculations to determine the ground state electronic configuration and detailed electronic structure parameters in V$X_3$. The trigonal distortion parameters $\Delta$$D_{3d}$ were determined to be -0.096 eV in VBr$_3$ and +0.070 eV in VI$_3$, revealing opposite signs of distortion in the two compounds and excellent agreement with experimental RIXS data. Cluster calculations confirm a high-spin V$^{3+}$ $(S = 1)$ configuration, with an $e'^2_g$ ground state in VBr$_3$ and an $e'^1_ga^1_{1g}$ ground state in VI$_3$, consistent with trigonal elongation and compression, respectively. These results reconcile prior discrepancies in the ground state electronic structure of V$X_3$, providing a robust experimental foundation for vanadium-based two-dimensional spintronic materials.

\end{abstract}
\maketitle

{\flushleft \textbf{Introduction}}

Two-dimensional (2D) materials have attracted intense research interest since the discovery of graphene, owing to their unique electronic and magnetic properties. In parallel, spintronics, an emerging field that exploits the electron’s spin degree of freedom in addition to its charge, has opened new pathways for next-generation electronic technologies, including nonvolatile memory, logic devices, and quantum information processing \cite{wolf2001spintronics,vzutic2004spintronics}. The discovery of intrinsic magnetism in low-dimensional materials has further accelerated this field, as 2D systems provide a versatile platform for controlling spin, orbital, and lattice degrees of freedom \cite{huang2017layer,gong2017discovery}. In particular, van der Waals magnets with tunable magnetic anisotropy and strong spin–orbit coupling (SOC) have emerged as promising candidates for spintronic applications \cite{gibertini2019magnetic}. Among them, transition metal (TM) trihalides with a honeycomb arrangement of the metal ions have emerged as particularly promising systems. Chromium trihalides Cr$X_3$ ($X$ = Cl, Br, and I) with $S = 3/2$ are well studied and are notable for sustaining long-range magnetic order down to the monolayer limit \cite{huang2017layer}. Their properties are strongly dependent on dimensionality, halogen chemistry, interlayer interactions, and temperature \cite{klein2018probing,jiang2018controlling,ahmad2020pressure,mishra2024magnetic}. 

Recent theoretical predictions and experiments have extended this interest to the vanadium trihalides VX$_3$ ($X =$ Br and I) with $S = 1$ \cite{kong2019crystal,he2016unusual,sant2023anisotropic,liu2020two}. The V$X_3$ undergo structural phase transitions from high-temperature monoclinic to low-temperature rhombohedral phase at 90 K (VBr$_3$) and 79 K (VI$_3$), similar to Cr$X_3$ \cite{kong2019vi3,lyu2022structural,tian2019ferromagnetic}. Subsequent magnetic phase transitions have been reported, where VBr$_3$ enters an antiferromagnetic (AFM) phase at 29.5 K, while VI$_3$ exhibits ferromagnetic (FM) order below 50 K \cite{kong2019vi3,lyu2022structural,hovanvcik2023robust}.

\begin{figure*}
    \centering
    {\includegraphics[width=1\textwidth]{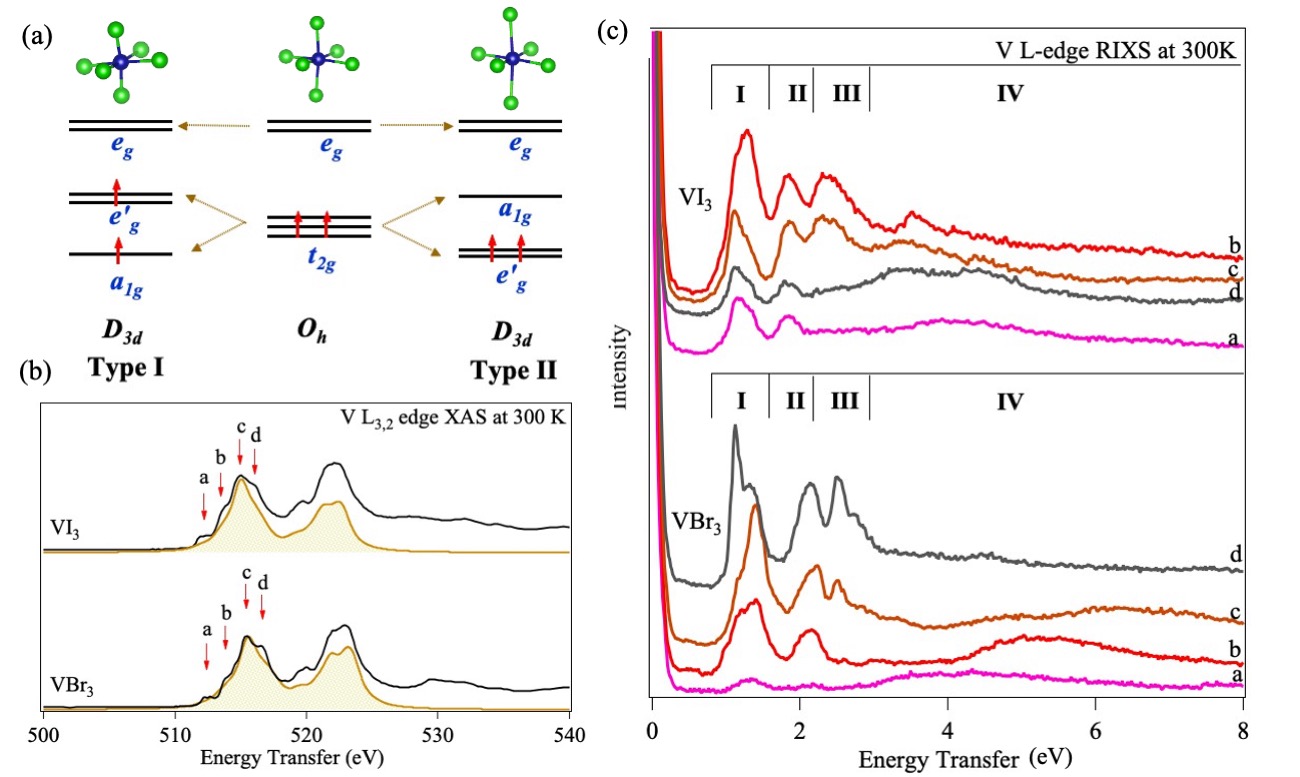}}
    \caption{Crystal field splitting and XAS/RIXS data in V$X_3$ (a) Crystal field splitting of the V $3d$ orbitals when the symmetry is lowered from $O_h$ to $D_{3d}$ symmetry. The Type I (II) configuration corresponds to the trigonal compression (elongation) in V$X_3$ as shown in the V$X_6$ cluster unit. (b) V $L_{3,2}$ edge XAS spectra in VI$_3$ and VBr$_3$ at 300 K, along with the simulated spectra indicated by the shaded region. (c) Excitation energy dependent RIXS measurements in V$X_3$ at 300 K. The excitation energies a, b, c, and d are as labeled in XAS spectra.}
    \label{fig:VX3_structure}
\end{figure*}

V$X_3$’s characteristic magnetic transitions and spin ordering mostly originate from the electronic structure of its V$^{3+}$ ions: V$X_3$ are layered van der Waals materials with a honeycomb lattice of V$^{3+}$ cations, each coordinated by six halogen anions in edge-sharing octahedra. In a perfect octahedral crystal field with $O_h$ symmetry, the V $3d$ orbitals split into two energy levels $t_{2g}$ and $e_g$. However, the partial filling of the $t_{2g}$ orbitals in V$X_3$ introduces Jahn-Teller (JT) distortions by splitting the $t_{2g}$ orbital into $a_{1g}$ and $e'_g$, lowering its symmetry from $O_h$ to $D_{3d}$ (see Figure \ref{fig:VX3_structure}(a)) \cite{georgescu2022trigonal,camerano2024symmetry,liu2020two}. This distortion can produce unquenched orbital moment and single-ion anisotropy via V$^{3+}$ SOC in contrast to Cr$X_3$, where anisotropy arises mainly from exchange interactions mediated by halogen orbitals \cite{pathiraja2025electronic}. 

Despite intensive study, the detailed crystal structure and electronic properties in V$X_3$ remain controversial. Literature propose two types of JT distortion in V$X_3$: (Type I) trigonal compression, yielding a ground state with half occupied $e_g'$ state and fully occupied $a_{1g}$ orbital ($e'^1_ga_{1g}^1$), and (Type II) trigonal elongation, where the $e_g'$ state is fully occupied ($e_g'^2$) \cite{sant2023anisotropic,yang2020vi,de2022influence}. These configurations correspond to distinct electronic behaviors: the Type I state typically results in metallicity, whereas the Type II state opens an insulating gap \cite{sant2023anisotropic,yang2020vi,kong2019vi3,he2016unusual,zhao2024multiple,son2019bulk,tian2019ferromagnetic}.

Theoretical studies have also suggested that including SOC in density functional theory calculations can induce additional splitting in the Type I $e_g'$ manifold for a reasonable Coulomb interaction $U$. This insulating state is argued to be stabilized by Mott correlation effects rather than JT distortion \cite{yang2020vi,liu2020two,tian2019ferromagnetic,son2019bulk,sant2023anisotropic}. Thus, while both theoretical and experimental evidence point to complex interplay among crystal-field, SOC, and correlation effects, a comprehensive experimental determination of the ground-state electronic configuration in V$X_3$ remains elusive. Understanding the electronic state near the Fermi level is crucial for unraveling the origin of magnetism in these compounds. Obtaining accurate energy scales is therefore a necessary step toward constructing robust theoretical models of the magnetic ground states in V$X_3$.

X-ray scattering techniques provide a powerful means of probing the electronic structure of TM-based structures and complexes. The development of synchrotron sources has enabled x-ray photon energies that directly match the relevant electronic transitions in many TMs. Resonant inelastic x-ray scattering (RIXS), in particular, is a $photon-in$ $photon-out$ spectroscopy that allows the study of the momentum, energy, and polarization changes of the scattered photon \cite{ament2011resonant}. In particular, the soft x-rays (0.1-2.0 keV) can excite TM $2p$ electrons into unoccupied valence orbitals at the TM $L-$edge, making RIXS particularly suited to probe $dd$ excitations and charge-transfer (CT) excitations in $3d$ or $4d$ TM complexes. \cite{ament2011resonant,pathiraja2025electronic}.

In this work, we report the first high-resolution RIXS measurements of VBr$_3$ and VI$_3$. The experimental spectra, combined with atomic multiplet calculations, provide direct evidence of the ground state electronic configuration and detailed electronic structure parameters in V$X_3$. The calculated energy level diagrams (ELDs) are compared with experimental RIXS spectra to optimize the energy scales. Our results identify an $e_g^2$ (trigonal elongation) ground state in VBr$_3$ and an $e_g'^1a_{1g}^1$ (trigonal compression) ground state in VI$_3$, consistent with prior theoretical predictions, providing a comprehensive experimental foundation for understanding the electronic and magnetic properties of V$X_3$ \cite{mudiyanselage2025determination}.

{\flushleft \textbf{Results}}

{\flushleft \textbf{V $L-$edge XAS and RIXS measurements}}

Figure \ref{fig:VX3_structure}(b) presents the V $L-$edge XAS data collected with total electron yield for the V$X_3$ compounds. The dipole allowed electronic transition ($2p^63d^2$ $\xrightarrow{}$ $2p^53d^3$) gives rise to two broad structures $L_3 (2p_{J=3/2}$ ($510-520$ eV) and $L_2 (2p_{J=1/2})$ ($520-530$ eV). The $L_3-$edge peak maxima occur at 515.5 eV for VBr$_3$ and 514.9 eV for VI$_3$, showing a clear energy shift as the halogen changes from Br to I, consistent with previous observations in Cr$X_3$ compounds \cite{pathiraja2025electronic,shao2021spectroscopic}. Additional XAS data can be found in Supplementary Figure S1.

Although the XAS provides valuable insight into key electronic excitations, its finite spectral broadening can complicate the precise determination of energy scales. On the other hand, it has been demonstrated that RIXS can provide more detailed information about electronic structure parameters with greater precision due to the sharpening effect in RIXS that bypasses core-hole lifetime broadening \cite{pathiraja2025electronic,ament2011resonant}. Figure \ref{fig:VX3_structure}(c) summarizes the V $L_3-$edge RIXS spectra in V$X_3$ collected at 300 K at different excitation energies marked in Figure\ref{fig:VX3_structure}(b) (Additional RIXS data can be found in Supplementary Figure S2).

The RIXS spectral features are identified in four main energy-transfer regions (I$-$IV): the regions I $-$ III are attributed to the inter-orbital $dd$ excitations, while region IV corresponds to broader charge-transfer (CT) excitations. The excitation energy dependent RIXS spectra showed only subtle peak shifts except for the CT features. However, their intensities varied clearly: the dominant peak in region I has high intensity at the $L_3$ Edge (excitation energy 'c'), whereas the intensity of the shoulder features increases as the excitation energy moves away from the $L_3$ peak. Notably, distinct behaviors are observed between the two compounds. In VI$_3$, the shoulder intensity is maximized at the pre-edge (excitation energy 'b'), while in VBr$_3$, the maximum occurs at the post-edge (excitation energy 'd'). This contrast suggests different excitation pathways in VBr$_3$ and VI$_3$ as a function of incident photon energy. The features associated with the $^3A_1$ and $^3E$ states, discussed in detail later, contribute differently in the two compounds, leading to these opposite trends in spectral weight distribution. 

{\flushleft \textbf{Temperature dependency in RIXS}}
\begin{figure} [!ht]
    \centering
    \includegraphics[width=0.5\textwidth]{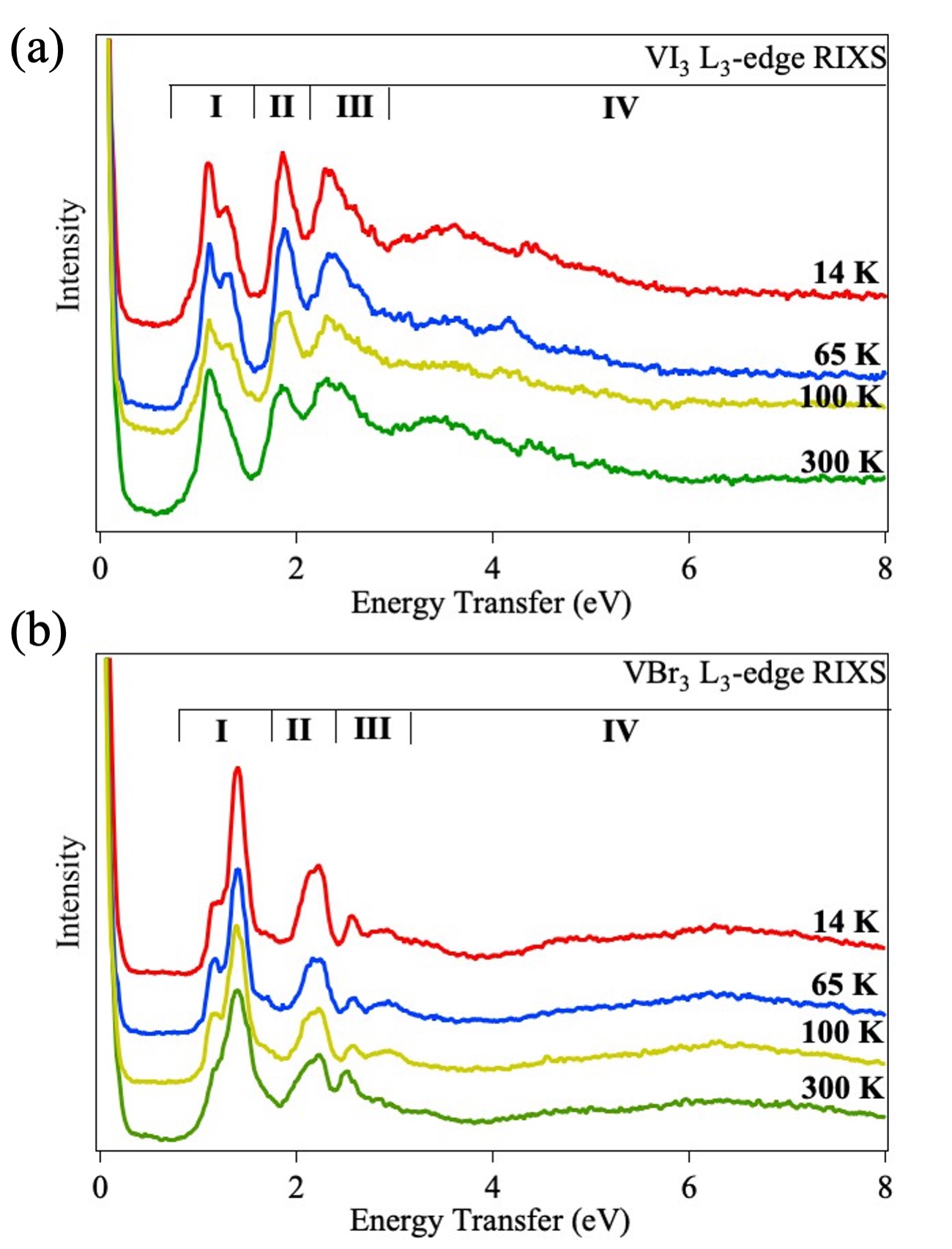}
    \caption{Temperature dependency in RIXS (a) Temperature dependent RIXS data in VBr$_3$ and (b) VI$_3$. All data were measured at the V $L_3-$edge with the incident beam at 90$^0$ to the sample surface. The data were collected at four temperatures: 14 K (red), 65 K (blue), 100 K (yellow), and 300 K (green). The energy transfer scales across all RIXS data are categorized into the same regions (I-IV).}
    \label{fig:VX3_RIXS}
\end{figure}

Figure \ref{fig:VX3_RIXS} presents the V $L_3-$edge RIXS spectra of V$X_3$, measured across different temperatures: 14, 65, 100, and 300 K. The temperature-dependent spectra correspond to different crystallographic phases: rhombohedral at 14 K and 65 K, and monoclinic at 100 K and 300 K. Both compounds are magnetically ordered at 14 K. Despite the different phases and magnetic transitions in V$X_3$, no substantial peak shifts are observed upon cooling, implying that the local environment around vanadium remains largely unchanged. Regarding the peak intensities, no clear intensity variation was observed across the different structural phases. However, the RIXS peak intensities of the shoulder features in regions I and III decrease as the temperature is lowered from 65K to 14K, while the main peak intensities remain unchanged. This means that the spin-allowed electronic transitions (triplet states) do not change, but the spin-forbidden electronic transitions (singlet states) have a direct impact when the temperature is lowered to the magnetic ordering temperature.

Comparing different compounds, VBr$_3$ and VI$_3$, in all RIXS data, clear spectral peak differences were observed across compounds. In region I, VBr$_3$ exhibits a shoulder feature on the low-energy side, whereas VI$_3$ shows a shoulder feature on the high-energy side. The primary peak maxima in this region occur at 1.4 eV (VBr$_3$) and 1.1 eV (VI$_3$), indicating a systematic shift towards lower energy as the halogen changes from Br to I, consistent with other transition metal halide systems \cite{pathiraja2025electronic}. In region III, VI$_3$ displays a single broader peak, while VBr$_3$ consists of two well-separated peaks with relatively small intensities. Furthermore, VBr$_3$ shows clearly resolved CT bands, whereas VI$_3$ exhibits significant overlap between $dd$ and CT excitations. In particular, at the $L_3-$edge, the CT bands are centered around 6.3 eV and 3.6 eV for VBr$_3$ and VI$_3$, respectively. This behavior reflects the higher binding energy of Br $4p$ orbitals, compared to I $5p$ orbitals, resulting in stronger hybridization between the V $3d$ and I $5p$ orbitals in VI$_3$. The increased spatial extent of the I $5p$ orbitals enhanced covalency, which suppresses the distinct CT peak intensity and shifts towards the onset of the intersite charge continuum. As a result, VI$_3$ exhibits broader and more overlapping spectral features compared to VBr$_3$. Similar trends have been reported in other halide systems such as Ru$X_3$ \cite{gretarsson2024j}.

\begin{figure*}
    \centering
    {\includegraphics[width=1 \textwidth]{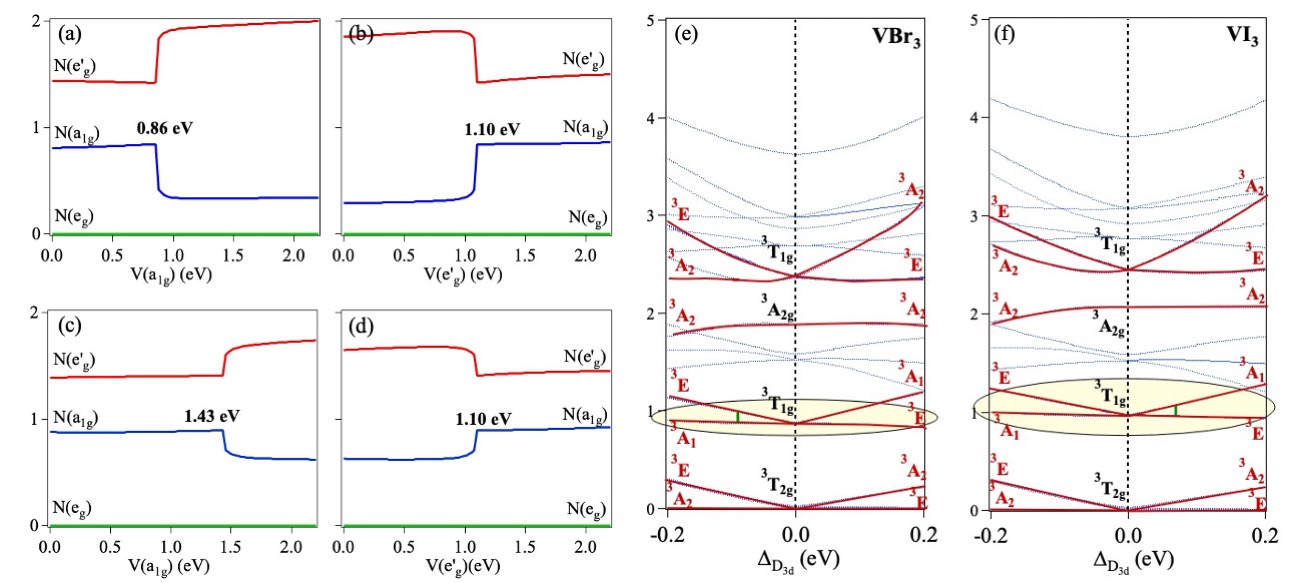}}
    \caption{ Electron occupancies and crystal field distortions in V$X_3$ (a$-$b) Calculated orbital occupations of the $e_g$ (green) $a_{1g}$ (blue) and $e'_g$ (red) states per V atom as function of the symmetry adapted potentials  $V(a_{1g})$ and $V(e'_g)$ in VBr$_3$ and (c$-$d) in VI$_3$ (e) Energy level diagram as a function of trigonal distortion parameter $\Delta$$D_{3d}$ in VBr$_3$ and (f) VI$_3$. At $\Delta$$D_{3d}$ = 0 eV, spectroscopic terms are labeled in $O_h$ symmetry, else they are labeled in $D_{3d}$ symmetry. The red and blue lines indicate the triplet and singlet states, respectively.}
    \label{fig:Quanty_parameters}
\end{figure*} 

{\flushleft \textbf{Atomic Multiplet calculations}}

All XAS and RIXS results indicate that the local environment in the metal V ion remains unchanged in all experimental geometries. Therefore, atomic multiplet calculations were performed using the $D_{3d}$ crystal field environment in the metal V ion to assign spectral features in XAS and RIXS spectra and to identify multiplet states. The calculations were performed using the quantum many-body script language $Quanty$ as explained in the Methods section \cite{haverkort2016quanty,haverkort2012multiplet}.

Initial starting parameters for the multiplet calculations were obtained from the references and the standard notations (The starting parameters of 10Dq = 1.10 eV, Racah B = 0.108 eV, and Racah C = 0.403 eV were selected for V$X_3$) \cite{sant2023anisotropic,yekta2021strength,yang2020vi,krzystek2004pseudooctahedral}. The on-site Coulomb repulsion $U_{dd}$ of 3.82 eV (VBr$_3$) and 2.91 eV (VI$_3$) was used in the simulations, while the ratio of $U_{dd}/U_{pd}$ was kept constant at 1.43 \cite{sant2023anisotropic}. The charge transfer energy ($\Delta$) was kept as 3.30 and 3.00 eV in VBr$_3$ and VI$_3$, following the same trend as CrX$_3$ \cite{sant2023anisotropic,shao2021spectroscopic}. The CT energy of 3.30 eV and 3.00 eV was used for VBr$_3$ and VI$_3$, respectively \cite{shao2021spectroscopic}.

{\flushleft \textbf{Calculated electron occupancy and ligand hybridization in V$X_3$}}

As the ligand hybridization affects the excited energy states significantly, the hybridization parameters were determined through a detailed calculation of electronic occupations in each $e_g$, $a_{1g}$, and $e'_g$ orbital. Figure \ref{fig:Quanty_parameters} (a$-$b) shows the calculated electron occupancy results as a function of the hybridization parameters $V(a_{1g})$ and $V(e'_g)$ in VBr$_3$ and (c$-$d) VI$_3$. In each panel, the number of electrons was counted in each orbital $e_g$ (green), $a_{1g}$ (blue), and $e'_g$ (red). There was no electron count in the $e_g$ orbital, and it remained constant. The electron count in $e'_g$ and $a_{1g}$ states is mainly related to the V $3d-$electrons modulated by the hybridization with the ligand (Br or I) ions, and they varied symmetrically as a function of $V(a_{1g}$ and $V(e'_g)$. 

From Figure \ref{fig:Quanty_parameters}, the $V(a_{1g})$ was determined to be 0.86 eV and 1.43 eV for VBr$_3$ and VI$_3$, respectively, while $V(e'_g)$ was 1.10 eV in both halide systems. The hopping integral $V(e_g)$ was kept constant at 1.7 eV in all calculations following Sant $et$ $al.$ \cite{sant2023anisotropic}. The ELDs also showed a strong dependence on the hopping integrals $V(a_{1g})$, $V(e'_g)$, and $V(e_g)$, and determining them from electron occupation was more accurate and consistent with other studies.

{\flushleft \textbf{Determination of trigonal distortion parameter $\Delta$$D_{3d}$}}
\label{Distortion_parameters}

Figure \ref{fig:Quanty_parameters}(e$-$f) shows the ELD calculated as a function of $\Delta$$D_{3d}$ in VBr$_3$ and VI$_3$, respectively. The dashed blue and solid red lines represent the spin-singlet and spin-triplet states, respectively. Under an $O_h$ crystal field with $e_g$ and $t_{2g}$ orbitals (when $\Delta$$D_{3d}$ = 0), the atomic triplet states split into the manifolds: $^3F$ $\xrightarrow{}$ $^3A_{2g}$, $^3T_{2g}$, and $^3T_{1g}(1)$, and $^3P$ $\xrightarrow{}$ $^3T_{1g}(2)$. As the JT distortion affects the $O_h$ symmetry to be distorted into the $D_{3d}$ symmetry, $t_{2g}$ orbitals are further splitted into $a_{1g}$ and $e'_g$ states, producing branching transitions such as; $^3T_{2g}$ $\xrightarrow{}$ $^3A_{1}$ x $^3E$ and $^3T_{1g}$ $\xrightarrow{}$ $^3A_{2}$ x $^3E$ \cite{dalal2017textbook}. The splitting of these multielectronic atomic states depends on the trigonal distortion parameter $\Delta$$D_{3d}$ \cite{pathiraja2025electronic}.

The following steps were taken in order to determine the crystal field trigonal distortion parameter $\Delta$$D_{3d}$. First, the region I in experimental RIXS data was fitted with five Gaussian peaks: two peaks for triplet states $^3E$ and $^3A_1$, and three peaks for singlet states $^1E$, $^1A_2$, and $^1E$ states. The experimental peak separation $|^3E-^3A_1|$ was obtained to be 0.108 eV and 0.131 eV in VBr$_3$ and VI$_3$, respectively. Comparing the determined peak separation with the ELD shown in Figure \ref{fig:Quanty_parameters}(e$-$f), the crystal field trigonal distortion parameters were determined to be -0.096 eV and 0.070 eV in VBr$_3$ and VI$_3$.

Since the VBr$_3$ experimental RIXS spectra showed a shoulder feature on the lower energy side in region I, the atomic multiplet ordering of $^3E > ^3A_1$ was chosen to align optical anisotropy and the RIXS intensity. Similarly, the ordering of $^3A_1 > ^3E$ was selected for VI$_3$, given the opposite shoulder feature arrangement in RIXS data.  The $^3E$ state's twofold symmetry aligns well with the anisotropy seen in the optical band structure, which is similar to other transition metal phosphorus trichalcogenides like NiPS$_4$ \cite{kim2024anisotropic}. The sign opposition between the determined VBr$_3$ and VI$_3$ suggests that there may be a sign opposition in the optical anisotropy, which warrants further discussion.

{\flushleft \textbf{Electronic structure parameters in VBr$_3$ and VI$_3$}}\label{ELD}

\begin{figure*}
    \centering
    \includegraphics[width=1\linewidth]{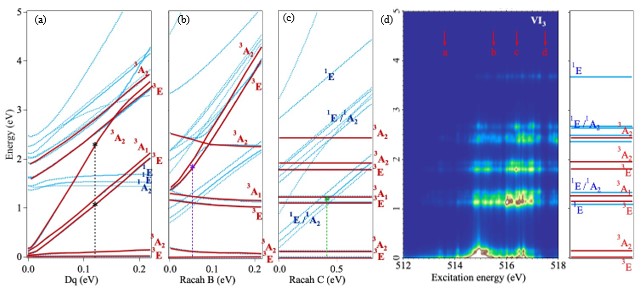}
    \caption{Simulation of RIXS spectra in VI$_3$ (a) Energy level diagrams as a function of crystal field $Dq$ (b) Racah $B$ and (c) Racah $C$ in VI$_3$. The triplet and singlet states in V$^{+3}$ metal are shown by solid red and dashed blue lines, respectively. ELDs are labeled with multiplet term symbols within $D_{3d}$ symmetry. The stars indicate the extracted parameter position by comparing with the experimental RIXS spectra. (d) Calculated RIXS map in VI$_3$. Final atomic multiplet energy states are shown next to the RIXS map.}
    \label{fig:VI3_ELD}
\end{figure*}

After determining all the preliminary parameters, the ELDs were computed and compared with experimental RIXS spectra to extract key electronic structure parameters, including crystal field splitting parameter $10Dq$, Racah $B$, and Racah $C$ in VI$_3$, as shown in Figures \ref{fig:VI3_ELD}(a$-$c) (see Supplementary Figure S4 for the ELDs in VBr$_3$).  

First, the $Dq$ was varied while keeping all other parameters constant, with Racah B and C set to the starting values of 0.108 eV and 0.403 eV, respectively. Figure \ref{fig:VI3_ELD}(a) shows the ELD depicting the evolution of multielectronic energy states as a function of $Dq$. Then, by comparing the energy positions of the first and third peak in the experimental RIXS spectra \cite{dalal2017textbook}, the crystal field splitting parameter $10Dq$ was determined considering the electronic transition  $^3A_{2}$ $\xrightarrow{}$ $^3E$ states (marked line). The extracted 10$Dq$ values are 1.370 eV (VBr$_3$) and 1.220 eV (VI$_3$), demonstrating a reduction in crystal field strength as the halogen changes from Br to I, consistent with trends in ligand-induced crystal fields \cite{pathiraja2025determination,tchougreeff2009nephelauxetic,pathiraja2025electronic}.

The Racah $B$ parameter, representing interelectronic repulsion, was determined to be 0.057 eV and 0.050 eV for VBr$_3$ and VI$_3$, respectively (see figure \ref{fig:VI3_ELD}(b) and Supplementary Figure S4(b)), indicating an increase in covalency from Br to I \cite{tchougreeff2009nephelauxetic,atanasov2012modern}. ELDs were also evaluated as a function of Racah parameter $C$, which predominantly influences singlet-state energies (see Figure \ref{fig:VI3_ELD}(c)). The triplet states appear nearly horizontal with increasing $C$, whereas the singlet states that are associated with spin-flip electronic transitions show clear dependency on Racah $C$. Optimal fits between the calculated and experimental RIXS spectra were obtained with Racah $C$ = 0.423 eV for both halide systems.

The exchange parameters $F^2_{pd}$, $G^1_{pd}$, and $G^3_{pd}$ were refined by fitting the XAS experimental data as shown in the yellow shaded line in Figure \ref{fig:VX3_structure}(b); as the RIXS process consists of an XAS process followed by a resonant x-ray emission. The vanadium $2p$ SOC of 4.65 eV is consistent with the observed energy gap between $L_3$ and $L_2$ edges, and in good agreement with literature \cite{lane2021two}. All the determined electronic structure parameters are summarized in the table \ref{tab: table energy scales}. 

\begin{table}
\caption{\label{tab: table energy scales}Summary of the energy scales calculated for V$X_3$ using multiplet ligand field theory calculations (all energies are in eV).}
    \begin{ruledtabular}
        \begin{tabular}{ccccc} 
            & \multicolumn{2}{|c|}{VBr$_3$} & \multicolumn{2}{|c|}{VI$_3$} \\ 
            & ($2p^63d^2$) & ($2p^53d^3$) & ($2p^63d^2$) & ($2p^53d^3$) \\ \hline
            $U_{dd}$ & 3.820 & 3.820 & 2.910 & 2.910 \\
            $U_{pd}$ & 5.463 & 5.463 & 4.161 & 4.161 \\
            $10Dq$ & 1.370 & 1.370 & 1.220 & 1.220 \\
            $\Delta$$D_{3d}$ & - 0.096 & - 0.096 & 0.070 & 0.070 \\
            Racah $B$ & 0.057 & 0.069 & 0.050 & 0.061 \\
            Racah $C$ & 0.423 & 0.459 & 0.423 & 0.459 \\
            $F^2_{dd}$ & 5.754 & 6.588 & 5.411 & 6.200 \\
            $F^4_{dd}$ & 5.330 & 5.778 & 5.330 & 5.778 \\
            $F^2_{pd}$ & - & 5.451 & - & 4.846 \\
            $G^1_{pd}$ & - & 3.733 & - & 3.075 \\
            $G^3_{3}$ & - & 2.496 & - & 2.496 \\
            SOC ($3d$) & 0.013 & 0.018 & 0.013 & 0.018 \\
            SOC ($2p$) & - & 4.650 & - & 4.650 \\
            $J_H$ & 0.792 & 0.792 & 0.767 & 0.767 \\
            E$(e_g)$ & 0.822 & 0.822 & 0.732 & 0.732 \\
            E$(a_{1g})$ & - 0.356 & - 0.356 & - 0.628 & - 0.628 \\
            E$_(e'_g)$ & - 0.644 & - 0.644 & - 0.418 & - 0.418 \\
        \end{tabular}
    \end{ruledtabular}
\end{table}

{\flushleft \textbf{Simulation of RIXS spectra}}

\begin{figure*}
    {\includegraphics[width=1 \textwidth]{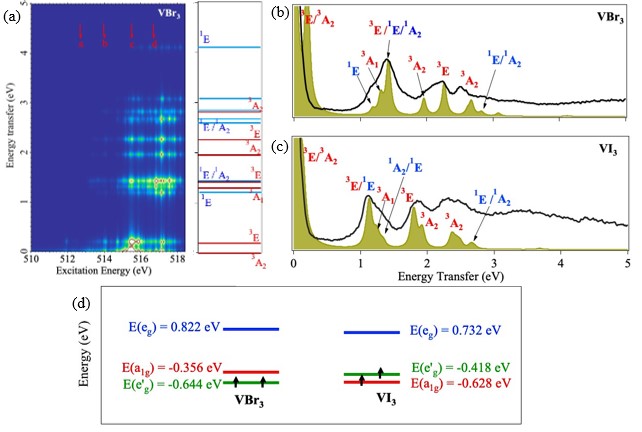}}
    \caption{ RIXS Simulations results and ground state electronic configurations (a) Calculated RIXS map in VBr$_3$. Final atomic multiplet energy states are shown next to the RIXS map. (c) The corresponding RIXS spectra calculated at the V $L_3-$edge (excitation energy 'c') in VBr$_3$ and (d) VI$_3$. The black line shows the experimental RIXS spectra, while the green-shaded spectra show the calculated RIXS spectra. The multielectronic atomic states, singlets and triplets, are labeled in blue and red, respectively. (e) Calculated crystal field splitting orbital energies in V$X_3$, showing their ground state electronic configurations.}
    \label{fig: RIXS_fitting_results_VX3}
\end{figure*}

After optimizing all the electronic structure parameters, the RIXS maps were calculated for VI$_3$ and VBr$_3$ as shown in Figure \ref{fig:VI3_ELD}(d) and \ref{fig: RIXS_fitting_results_VX3}(a), respectively, along with the atomic multiplet energy levels. The calculations reported here show results at 300 K, and no significant temperature dependence was observed in the RIXS spectral features at low temperatures. We find excellent agreement between the experimental and simulation data, including the energies of all Raman-like excitations in the RIXS profile and their incident-energy dependence.

The RIXS line spectra at the V $L_3-$edge (see marked excitation energy 'c' in RIXS maps) in VBr$_3$ and VI$_3$ are shown in Figure \ref{fig: RIXS_fitting_results_VX3}(b$-$c). The black line shows the experimental RIXS spectra, while the yellow shaded spectra show the calculated RIXS spectra. The spectral features have been labeled with spin-allowed triplet states (red) and spin-forbidden singlet states (blue), respectively. The triplet states $^3A_1$ and $^3E$ showed an ordering of $^3A_1>^3E$ and $^3A_2>^3E$ in VBr$_3$, while it is switched to $^3E>^3A_1$ and $^3E>^3A_2$ in VI$_3$. The atomic multiplet calculations justify that this is consistent with the sign opposition in the trigonal distortion parameter $\Delta$$D_{3d}$ between VBr$_3$ and VI$_3$. Note that the CT excitations are not captured in our calculations. 

The temperature dependency in experimental RIXS spectral features in Figure \ref{fig:VX3_RIXS} can be explained using the spin-forbidden singlet states $^1A$ and $^1E$. These singlet states are highly dependent on SOC, and its inclusion was required to produce the spin-forbidden singlet states. When SOC = 0, only the triplet states were observed. As the temperature decreased, we observed a clear variation in the intensity of the spectral features. As the temperature is lowered, the spins become more stable and oriented and produce a strong SOC, the RIXS shoulder peak intensities have diminished at 14K. Therefore, the change of RIXS spectral features intensity as the temperature decreases (see Figure \ref{fig:VX3_RIXS}) can be directly linked with the magnetic ordering in V$X_3$. Similar behavior was observed in other TM complexes, ruby, Cr$X_3$, and Vanadium oxides \cite{hunault2018direct,pathiraja2025electronic,zheng2011theoretical}.

{\flushleft \textbf{Ground state electronic configuration in V$X_3$}}

Through this study, we experimentally determined the ground state electronic configuration of V$X_3$. The energies of the $e_g$, $a_{1g}$, and $e_g'$ states are summarized in Table \ref{tab: table energy scales} and are graphically shown in Figure \ref{fig: RIXS_fitting_results_VX3}(d) (Please see Supplementary Figure S3 for the crystal field splitting in metal V$^{3+}$ ion). The calculations indicate that VBr$_3$ has the lowest state of $e'_g$, indicating a ground state of $e'^2_g$ (Type II in Figure \ref{fig:VX3_structure}(a)) \cite{camerano2024symmetry}. The full occupancy of the $e'_g$ orbitals and empty $a_{1g}$ orbitals introduces a clear energy gap between the two orbitals, resulting in an insulating state. First principal calculations also suggest an $e_g'^2$ ground state in VBr$_3$, consistent with its reported semiconducting behavior, showing excellent agreement with our findings \cite{liu2020two,kong2019crystal}. The energy gap between the E$(e_g)$ and E$(a_{1g})$ is determined to be 1.178 eV and the energy gap between E$(a_{1g})$ and E$(e'_g)$, which was determined to be 0.288 eV, is responsible for opening a narrow bandgap with a reasonable Coulomb interaction $U$, resulting in the insulating state in VBr$_3$, showing good agreement with the literature \cite{liu2020two}.

In contrast to VBr$_3$, we determined the lowest energy of VI$_3$ as $a_{1g}$, indicating a ground state electronic configuration of $e'^1_ga_{1g}^1$ (Type I in Figure \ref{fig:VX3_structure}(a)). Similar ground state configurations have been reported by Yang $et$ $al.$ and Sant $et$ $al.$ \cite{sant2023anisotropic,yang2020vi}. Although this ground state with partial occupancy in $e'_g$ orbital leads to a metallic state in nature, it is also reported that the V $3d$ SOC plays a significant role, and with a reasonable on-site Coulomb repulsion $U$, a splitting of the two $e'_g$ orbitals can be observed, introducing an insulating state in VI$_3$ \cite{yang2020vi}. In particular, Zhao $et$ $al.$ reports bandgap opening of 12.7 meV upon including the SOC in VI$_3$, which is consistent with the V $3d$ SOC from our multiplet calculations (13 meV) \cite{zhao2024multiple}. Another report indicates a bandgap of 38.7 meV, and this elevation in SOC was due to the strong I $5p$-V $3d$ hybridization along with the on-site Coulomb repulsion $U$ \cite{tian2019ferromagnetic,son2019bulk}.

Our calculations suggest that the primary factor governing the distinct ground state configurations in V$X_3$ is the trigonal distortion parameter $\Delta$$D_{3d}$. The ground state configuration is highly sensitive to both the sign and the magnitude of the $\Delta$$D_{3d}$: a negative $\Delta$$D_{3d}$ in VBr$_3$ leads to an orbital doublet ground state ($e_g'^2$), whereas a positive $\Delta$$D_{3d}$ in VI$_3$ stabilizes an orbital singlet ground state ($e'^1_ga_{1g}^1$). In the strong crystal field basis, the doublet state corresponds to the $d_{xy}$,$d_{yz}$ orbitals, while the singlet arises from the $d_{z^2}$ orbital. Therefore, the hybridization with ligands results in a trigonal elongation for VBr$_3$ and a trigonal compression for VI$_3$, as graphically shown in Figure \ref{fig:VX3_structure}(a).

{\flushleft \textbf{Discussion}}

We have reported detailed electronic structure parameters and ground state electronic configuration in V$X_3$ ($X$ =Br and I) using high-resolution RIXS measurements with atomic multiplet calculations. From the cluster calculations, a spin moment of 2 $\mu_B$ per V atom was obtained, indicating a high-spin state in V$^{3+}$ with $S=1$. This result is in excellent agreement with the expected value, and aligns well with theoretical predictions of approximately $2.2-2.6$ $\mu_B$ per V \cite{kong2019vi3,tian2019ferromagnetic}. The Hund’s exchange coupling $J_H = (F^2_{dd} + F^4_{dd})/14$ was determined to be 0.792 (VBr$_3$) and 0.767 eV (VI$_3$), confirming the high-spin state in both compounds, and consistent with the relation 2$J_H$ $<$ 10$Dq$ \cite{tomiyasu2017coulomb}. Literature reports a $J_H$ of 0.8-0.9 eV in both V$X_3$, showing excellent agreement with our results \cite{hovancik2023large,yang2020vi,liu2020two,anisimov1997first}. 

The high-resolution RIXS spectra with multiple spectral features enabled a precise determination of the trigonal distortion parameter $\Delta$$D_{3d}$. The obtained $\Delta$$D_{3d}$ in VI$_3$ (0.07 eV) is consistent with previous reports, while our study provides the first experimental determination of $\Delta$$D_{3d}$ in VBr$_3$ (-0.096 eV) \cite{lane2021two,hovancik2023large}. These findings confirm the critical role of the trigonal distortion in defining the electronic ground states in V$X_3$. Although our results establish strong experimental evidence for the distinct electronic configurations, further investigations are warranted to quantitatively assess the SOC-induced splitting within the $e'_g$ manifold and its contribution to the Mott insulating behavior in VI$_3$.

In conclusion, we have investigated the electronic structure and ground state configurations of vanadium trihalides V$X_3$ (X = Br and I) using high-resolution RIXS measurements combined with ligand field multiplet calculations, and presented critical insights into the electronic and orbital ground states of V$X_3$, which are directly relevant to spintronic applications. In two-dimensional magnetic systems, spintronic functionality, such as spin transport, magnetic switching, and anisotropic behavior, is strongly governed by the interplay among crystal field splitting, SOC, and orbital occupation. Our results demonstrate that trigonal distortion controls the orbital ground state, leading to distinct electronic configurations in VBr$_3$ and VI$_3$ ($e'^2_g$ ground state in VBr$_3$ and an $e'^1_ga^1_{1g}$ ground state in VI$_3$). In particular, the presence of an orbital singlet versus doublet state influences magnetic anisotropy and spin dynamics, key parameters for stable spintronic device operation. Furthermore, the SOC-driven Mott-insulating behavior observed in VI$_3$ highlights the potential to tune electronic phases through orbital and lattice engineering. The experimentally extracted parameters are in excellent agreement with theoretical and optical studies, providing the most complete experimental picture of the low-energy electronic structure in V$X_3$ to date. The precise determination of crystal field parameters and energy scales reported here establishes a quantitative framework for designing and optimizing 2D van der Waals magnets with controllable spin and orbital degrees of freedom, enabling next-generation low-power, quantum, and spin-based electronic devices.

{\flushleft \textbf{Method}}
{\flushleft \textbf{Resonant Inelastic X-ray Scattering experiment}}

V$X_3$ single crystals were obtained from HQ Graphene (Groningen, Netherlands). Due to the high hygroscopicity and oxygen sensitivity of the V$X_3$ samples, they were stored and handled under an inert atmosphere (Ar) in a glovebox to minimize air exposure. The samples were subjected to scotch tape exfoliation before being transferred to the experimental chamber to ensure a clean surface. The experimental chamber was maintained in dark conditions and at a high vacuum (5 × 10$^{-9}$ Torr) during the measurements.

The preliminary V$X_3$ $L_{2,3}$ XAS measurements and RIXS measurements were obtained at the qRIXS end station, beamline 8.0.1 at the Advanced Light Source, Lawrence Berkeley National Laboratory, with a resolution of 0.3 eV \cite{chuang2022momentum}. High-resolution XAS and RIXS data were collected at the PEAXIS beamline at BESSY II, Germany, with a resolution of 80 meV \cite{schulz2020characterization}. 

The data were collected at four temperatures: 14, 65, 100, and 300 K. The data reported in the main text were measured with the incident photon beam at $90^0$ relative to the sample surface, and the RIXS spectrometer at $135^0$ (backscattering geometry). At all experimental conditions, the x-ray photon beam was horizontally polarized to the scattering plane ($\pi$-polarization). The crystalline $c$-axis was aligned within the scattering plane. A closed cycled He cryostat flow was used in the low temperature RIXS measurements.

{\flushleft \textbf{V$X_6$ Cluster calculations}}

All XAS and RIXS results indicate that the local environment in the metal V ion remains unchanged in all experimental geometries. Therefore, atomic multiplet calculations were performed using the $D_{3d}$ crystal field environment in the metal V ion to assign spectral features in XAS and RIXS spectra and to identify multiplet states. The calculations were performed using the quantum many-body script language Quanty \cite{haverkort2016quanty,haverkort2012multiplet}. 

The large V-$X$ covalency necessitates the consideration of CT effects, including the $3d^2$ and $3d^3L$ electronic configurations in the calculations, where $L$ corresponds to a hole in the ligand orbitals (I $5p$). The ligand field Hamiltonian can be expressed as,
\begin{equation}
    H = H_{Coulomb}+H_{SOC}+H_{CFT}+H_{LMCT}+H_{Exc.}
\end{equation}
including  the  interorbital Coulomb interaction effect ($H_{Coulomb}$), crystal field effect ($H_{CFT}$), spin-orbit coupling ($H_{SOC}$), the ligand to metal charge transfer effect ($H_{LMCT}$), and exchange Heisenberg interaction effect ($H_{Exc.}$) \cite{haverkort2016quanty,haverkort2012multiplet,atanasov2004dft,woolley1987ligand}. 

Quanty uses the second quantization to solve the electronic Hamiltonian and to calculate the RIXS spectra and ELDs. The relevant electronic configurations of $2p^63d^2$ and $2p^53d^3$ in the ground state and the excited state resulted in the atomic multiplets $^3F < ^1D < ^3P < ^1G < ^1S$. These multiplets can be described by $3d-3d$ Coulomb and $2p-3d$ exchange interactions parameterized in Slater-Condon integrals $F^k_{dd}$, $F^k_{pd}$ (Coulomb), and $G^k_{pd}$ (Exchange) for Hatree-Fock calculations \cite{haverkort2016quanty,hunault2018direct}. They can be related to the Racah $B$ and $C$ by,
\begin{equation}
\begin{split}
     \text{Racah \hspace{0.2cm}} B & = (9F^2_{dd} - 5F^4_{dd})/441 \\
     \text{Racah \hspace{0.2cm}} C & = 5F^4_{dd}/63
\end{split} 
\end{equation}
to account for the ion covalency \cite{racah1942theory}.

{\flushleft \textbf{acknowledgments}}
This research was conducted at PEAXIS RIXS beamline, BESSY II, Germany, operated by the Helmholtz-Zentrum Berlin für Materialien und Energie, and qRIXS 8.0.1 beamline, Advanced Light Source (ALS), which is a DOE Office of Science User Facility under contract no. DE-AC02-05CH11231. The Welch Foundation (grant number: E-0001) and the Texas Center for Superconductivity (TCSUH) supported work at the University of Houston. Part of this work was supported by the U.S. DOE, BES, under Award No. DE-SC0024332. The authors acknowledge support from the U.S. Air Force Office of Scientific Research and Clarkson Aerospace Corp. under Award FA9550-21-1-0460. Y.C.S. acknowledges the financial support from the National Science and Technology Council (NSTC) in Taiwan under grant numbers 113-2112-M-213-025-MY3. Special thanks to the eXn group members at the University of Houston.

\section{Supplementary}
\section{S1: Angular temperature dependency in XAS}

\begin{figure*}
    \centering
    {\includegraphics[width=1 \textwidth]{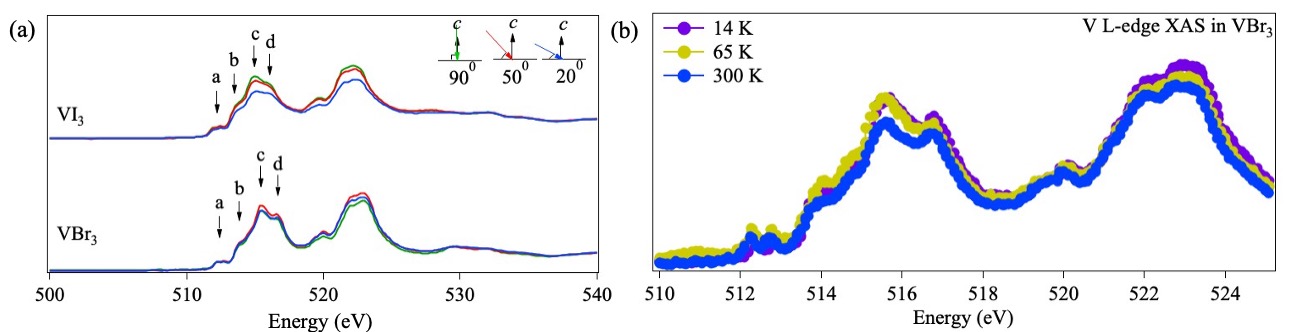}}
    \caption{ ((a) V $L$ edge XAS data in VBr$_3$ and VI$_3$ at 300 K at different x-ray incident geometries. The blue, red, and green lines indicate the XAS data collected at incident angles of $20^0$, $50^0$, and $90^0$ (see the top-right schematic diagram). (b) Temperature dependency in VBr$_3$, measured at the $90^0$ incidence geometry.}
    \label{fig:XAS_VX3}
\end{figure*}

Here, we report the angle dependent and temperature dependent X-ray absorption (XAS) data. Figure \ref{fig:XAS_VX3}(a) shows the XAS data collected at room temperature at three different incident photon beam geometries: grazing incidence $20^0$ (blue), $50^0$ (red), and normal incidence $90^0$ (green). No significant peak shifting was observed across different incident angles; however, an intensity variation was observed. At the $L_3-$ edge, the normal incidence shows the highest intensity, and the grazing incidence exhibits the lowest intensity. Decreasing the incident angle to a "grazing" condition significantly reduces x-ray penetration depth, making the XAS signal much more sensitive to the surface or near-surface region rather than the bulk. Therefore, at the $\pi$ polarization, the grazing incidence XAS primarily probes the surface region, whereas the normal incidence XAS probes deeper into the bulk. A pronounced variation in intensity with angle was observed in VI$_3$, consistent with the depth-dependent sensitivity described above. Additionally, there was no significant O $K$ edge intensity observed in the 530–540 eV region, indicating no sign of oxide contamination. The excellent agreement between bulk sensitive RIXS (normal incidence) and multiplet calculations further suggests that the observed spectra originate from intact vanadium halide layers rather than degraded surfaces.

Figure \ref{fig:XAS_VX3}(b) shows the temperature dependent XAS in VBr$_3$ (similar results were observed for VI$_3$). The temperature-dependent V $L-$edge XAS spectra measured at 14 K, 65 K, and 300 K,  spanning over the structural and magnetic phase transition temperatures, show no significant energy shift, while only slight intensity variations could be observed. The 300 K XAS spectrum shows a distinct intensity redistribution compared to the 14 K and 65 K spectra. This indicates a clear relationship to the different structures at given temperatures, as V$X_3$ is in the monoclinic structure at 14 K and 65 K, and the trigonal structure at 300 K. The absence of significant energy changes at different temperatures indicates that the local environment surrounding V near the sample surface has only undergone subtle changes.

\section{S2: Additional RIXS data}

\begin{figure*}
    \centering
    \includegraphics[width=1\textwidth]{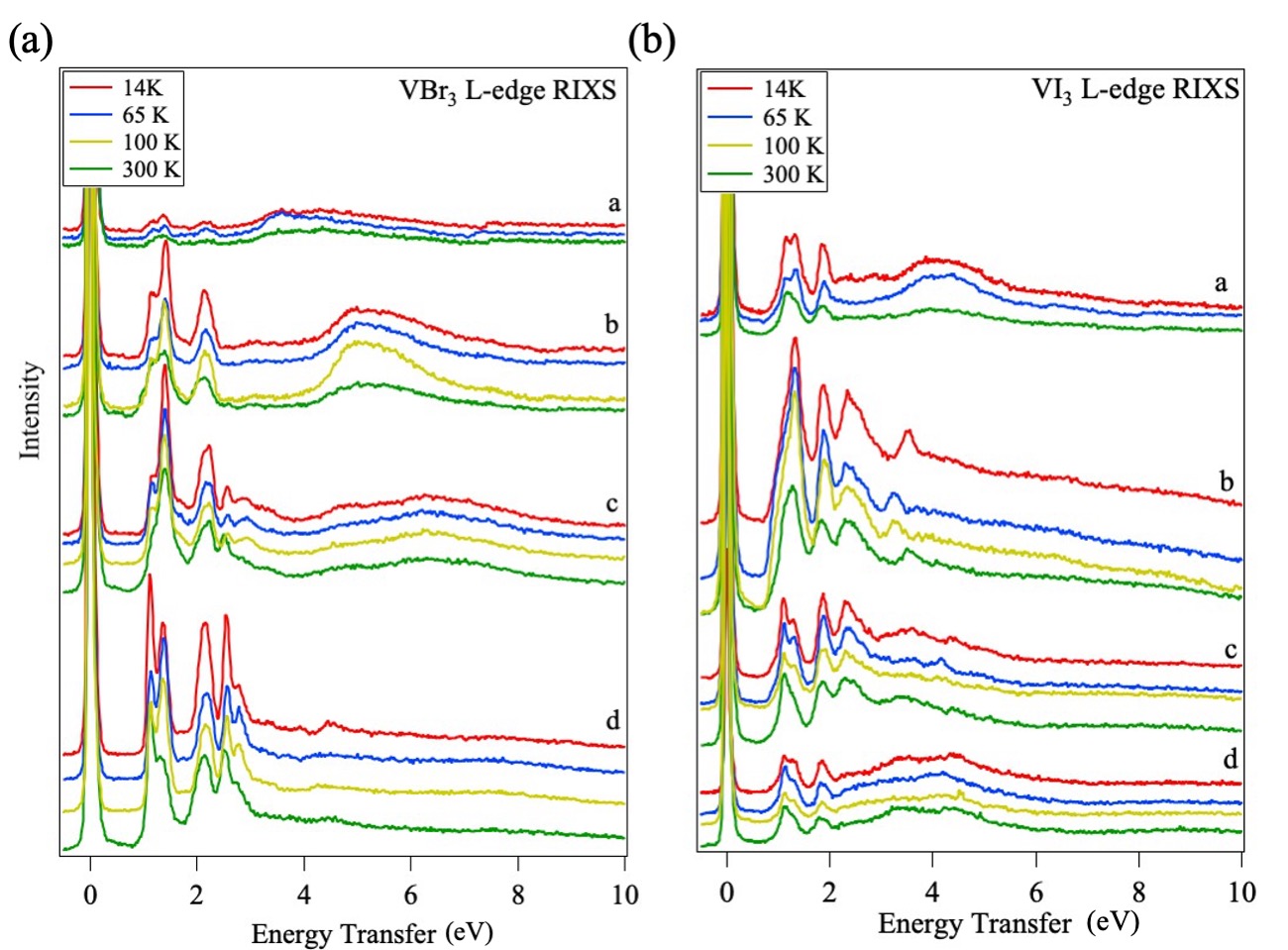}
    \caption{(a) Additional temperature dependence RIXS measurements in VBr$_3$ and (b) VI$_3$ at normal incidence. The data was collected at four temperatures 14 K, 65 K, 100 K, and  300 K, and four different excitation energies, a, b, c, and d, as labeled in Figure 2. }
    \label{fig:VX3_additional_RIXS}
\end{figure*}

Here, we report additional incident-energy dependent and temperature dependent resonant inelastic X-ray scattering (RIXS) data. These results are performed in the same way as the measurements shown in the main text. The experimental configuration is fixed at a scattering angle $2\theta = 135^0$. All the data were collected at normal incidence ($k_{in}$ is at $90^0$ degrees with respect to the sample surface). Different colors, red, blue, yellow, and green lines indicate the RIXS data collected at different temperatures, 14, 65, 100, and 300 K, respectively. Similar behavior was observed, with only slight intensity variations in the RIXS spectral features as the temperature changed. Incident-energy dependence also shows consistent behaviors at all measured temperatures.

\section{S3: Crystal field splitting in V$X_3$}

The crystal field environment in V$X_3$ was introduced using $D_{3d}$ symmetry. Under an $O_h$ crystal field, the V $5d$ orbitals are split into two energy levels $e_g$ and $t_{2g}$. The JT distortion under $D_{3d}$ symmetry further splits the $t_{2g}$ orbitals into $a_{1g}$ and $e'_g$ states. Therefore, we can express the energy of each orbital in $D_{3d}$ crystal field environment as,
\begin{equation}
E(e_g) = 6Dq \\
\end{equation}
\begin{equation}
E(a_{1g}) = -4Dq -2\Delta_{D_{3d}} \\
\end{equation}
\begin{equation}
E(e'_g) = -4Dq + \Delta_{D_{3d}} 
\label{eq: d_level_energies}
\end{equation}
where $\Delta$$D_{3d}$ is the trigonal distortion parameter. 

\begin{figure}[!h]
    \centering
    \includegraphics[width=1\linewidth]{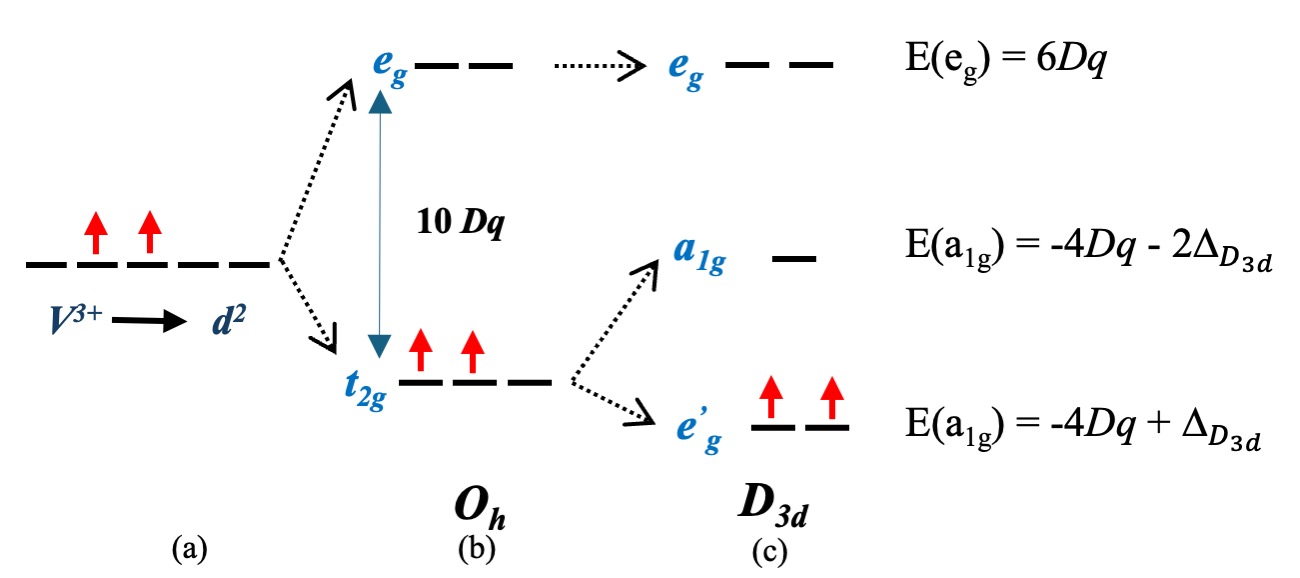}
    \caption{(a) valence $d$ orbital electron distribution in Cr$^{3+}$ metal ion. (b) Lifting of degeneracy of the $d^{2+}$ spectroscopic term (free $F^{4+}$ ion) due to $O_h$ symmetry. This configuration’s 5 $d$ orbitals are divided into two energy levels $t_{2g}$ and $e_g$. (c) Lifting of the degeneracy of the $d^{2+}$ electrons due to $D_{3d}$ symmetry. The five $d$ orbitals are divided into one $a_{1g}$ state and two $e_g$ states within the $D_{3d}$ symmetry.}
    \label{fig: d_level_splitting}
\end{figure}

\section{S4: Energy level diagrams for VBr$_3$}

\begin{figure}[!h]
    \centering
    {\includegraphics[width=0.5 \textwidth]{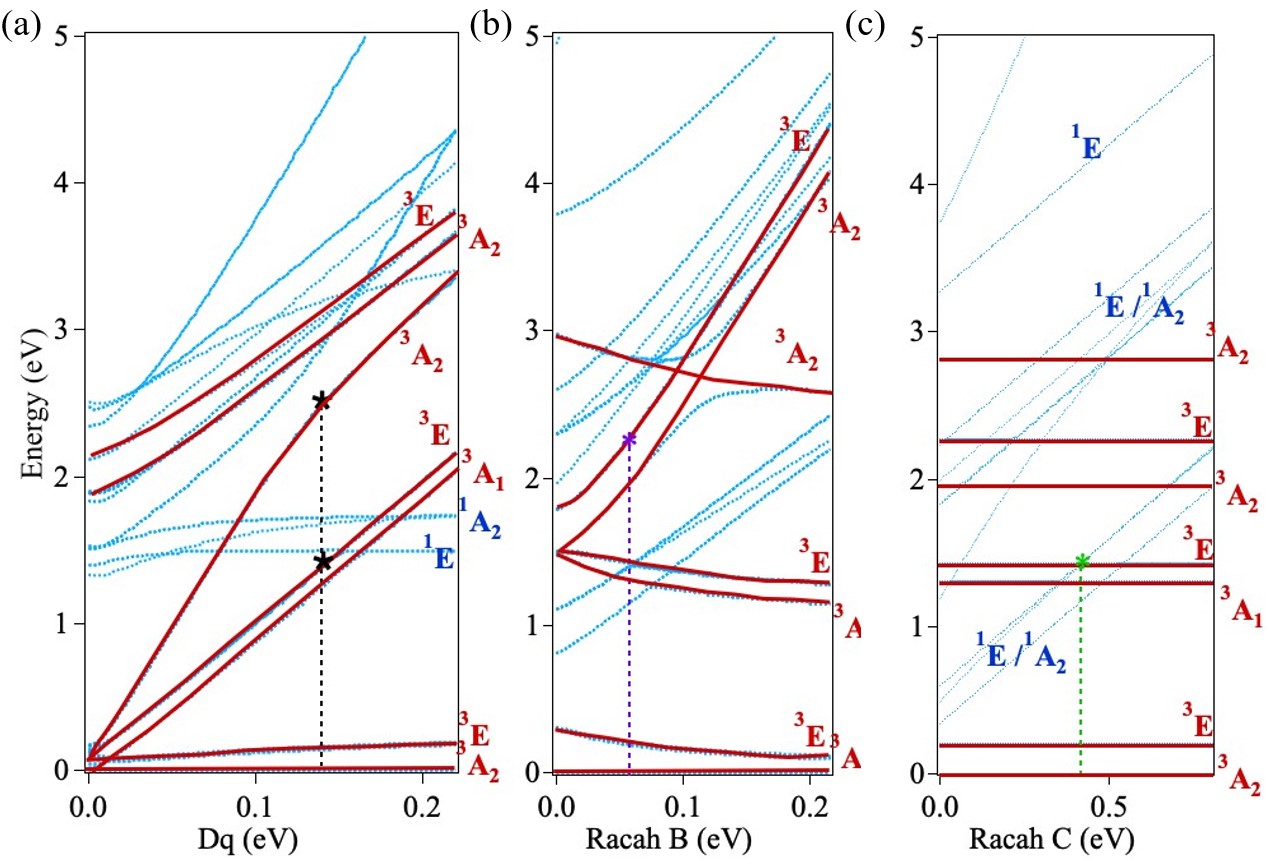}}
    \caption[Energy level diagrams in VBr$_3$ as a function of $Dq$, Racah $B$, and Racah $C$ ]{ (a) Energy level diagrams as a function of crystal field $Dq$, (b) Racah $B$, and (c) Racah $C$ in VBr$_3$. The triplet and singlet states in V$^{+3}$ metal are shown by solid red and dashed blue lines, respectively. The black, purple, and green dashed lines indicate the extracted energy scales.}
    \label{fig: ELD in VBr3}
\end{figure} 

Figure \ref{fig: ELD in VBr3} shows the ELDs for determining the crystal field splitting 10$Dq$, Racah $B$, and Racah $C$ in VBr$_3$. To obtain the energy parameters for simulating the VBr$_3$ RIXS spectra, the same approach was taken as VI$_3$ in the main text. 

First, the ELD was calculated as a function of $Dq$ with fixed initial Racah $B$ (0.108 eV) and Racah $C$ (0.403 eV), as shown in Figure \ref{fig: ELD in VBr3}(a). Then the first and third peak of the experimental RIXS spectra were compared with the $^3A_{2}$ and $^3E$ lines to consider the electronic transition  $^3A_{2}$ $\xrightarrow{}$ $^3E$, and the $10Dq$ value was determined to be 1.370 eV. 

Next, by keeping the determined $10Dq$ value constant with the initial Racah $C$, the ELD was calculated by varying Racah $B$ as shown in Figure \ref{fig: ELD in VBr3}(b) and determined to be 0.057 eV, following the same approach as VI$_3$, as mentioned in the main text. Finally, the newly determined $10Dq$ and Racah $B$ were kept constant, and the ELD was calculated by varying the Racah $C$ as shown in Figure \ref{fig: ELD in VBr3}(c). The singlet states were used to determine the Racah $C$ value of 0.423 eV.

\bibliographystyle{unsrt}
\bibliography{revision_2}

{\flushleft \textbf{Data availability}}
The raw data supporting the findings of this study are available under the following DOI: https://doi.org/10.18738/T8/JCVLYZ. 

{\flushleft \textbf{Code availability}}
The Quanty scripts used for the simulations are available from the corresponding author upon reasonable request.

{\flushleft \textbf{Author Contributions}}
B.F. supervised the project and acquired funding support. C.P. conducted the XAS/RIXS experiments with the support of Y.C., D.W., and C.S. C.P. performed data analysis, atomic multiplet calculations in discussion with Y.C., Y.S., and B.F. C.P. wrote the paper, incorporating input from all co-authors.

{\flushleft \textbf{Competing interests}}
The authors declare no competing financial or non-financial interests.

\end{document}